\documentclass[conference]{IEEEtran}

\usepackage{graphicx}
\usepackage{cite}
\usepackage{graphicx}%
\usepackage{multirow}%
\usepackage{amsmath,amssymb,amsfonts}%
\usepackage{amsthm}%
\usepackage{mathrsfs}%
\usepackage{xcolor}%
\usepackage{textcomp}%
\usepackage{manyfoot}%
\usepackage{booktabs}%
\usepackage{algorithm}%
\usepackage{algorithmicx}%
\usepackage{algpseudocode}%
\usepackage{listings}%
\usepackage{caption}
\usepackage{subcaption}
\usepackage[normalem]{ulem}

\begin{document}

\title{A lightweight Convolutional Neural Network based on U shape structure and Attention Mechanism for Anterior Mediastinum Segmentation}

\author{
    \IEEEauthorblockN{
        \begin{tabular}{c}
            Sina Soleimani-Fard\textsuperscript{1}, Won Gi Jeong\textsuperscript{2}, Francis Ferri Ripalda\textsuperscript{1}, \\
            Hasti Sasani\textsuperscript{3}, Younhee Choi\textsuperscript{1}, S Deiva\textsuperscript{4},Gong Yong Jin\textsuperscript{5}, Seok-bum Ko\textsuperscript{1}
        \end{tabular}
    }
    \IEEEauthorblockA{
        \begin{tabular}{c}
            \textsuperscript{1}Department of Electrical and Computer Engineering, University of Saskatchewan, 57 Campus Drive, \\
            Saskatoon, SK, Canada\\
            Email: \{Vuy825, kzn518, younhee.choi, seokbum.ko\}@usask.ca
        \end{tabular}
    }
    \IEEEauthorblockA{
        \begin{tabular}{c}
            \textsuperscript{2}Department of Radiology, Chonnam National University Hwasun Hospital, Hwasun, South Korea\\
            Email: wgjung86@naver.com
        \end{tabular}
    }
    \IEEEauthorblockA{
        \begin{tabular}{c}
            \textsuperscript{3}Department of Electrical Engineering, Tarbiat Modares University, Al Ahmad Street, Jalal No. 7, Tehran, Iran\\
            Email: Hastisasani@modares.ac.ir
        \end{tabular}
    }
        \IEEEauthorblockA{
        \begin{tabular}{c}
            \textsuperscript{4}Department of Electrical Engineering, National Institute of Technology, Trichy, India.\\
            Email: deiva@nitt.edu
        \end{tabular}
    }
    \IEEEauthorblockA{
        \begin{tabular}{c}
            \textsuperscript{5}Department of Radiology, Research Institute of Clinical Medicine of Jeonbuk National University, \\
            Biomedical Research Institute of Jeonbuk National University Hospital, Jeonbuk National University Medical School, \\
            20 Geonji-ro, Jeonju City, South Korea\\
            Email: gyjin@jbnu.ac.kr
        \end{tabular}
    }

}

\maketitle

\begingroup
\renewcommand\thefootnote{\fnsymbol{footnote}}
\footnotetext[1]{Sina Soleimani-Fard is the first author. Won Gi Jeong is the co-first author. Seok-bum Ko is the corresponding author. Gong Yong Jin is a co-corresponding author.}
\endgroup

\begin{abstract}
To automatically detect Anterior Mediastinum Lesions (AMLs) in the Anterior Mediastinum (AM), the primary requirement will be an automatic segmentation model specifically designed for the AM. The prevalence of AML is extremely low, making it challenging to conduct screening research similar to lung cancer screening. Retrospectively reviewing chest CT scans over a specific period to investigate the prevalence of AML requires substantial time. 
Therefore, developing an Artificial Intelligence (AI) model to find location of AM helps radiologist to enhance their ability to manage workloads and improve diagnostic accuracy for AMLs.
In this paper, we introduce a U-shaped structure network to segment AM. Two attention mechanisms were used for maintaining long-range dependencies and localization. In order to have the potential of Multi-Head Self-Attention (MHSA) and a lightweight network, we designed a parallel MHSA named Wide-MHSA (W-MHSA). Maintaining long-range dependencies is crucial for segmentation when we upsample feature maps. Therefore, we designed a Dilated Depth-Wise Parallel Path connection (DDWPP) for this purpose. 
In order to design a lightweight architecture, we introduced an expanding convolution block and combine it with the proposed W-MHSA for feature extraction in the encoder part of the proposed U-shaped network. 
The proposed network was trained on 2775 AM cases, which obtained an average Dice Similarity Coefficient (DSC) of 87.83\%, mean Intersection over Union (IoU) of 79.16\%, and Sensitivity of 89.60\%.
Our proposed architecture exhibited superior segmentation performance compared to the most advanced segmentation networks, such as Trans\_Unet, Attention\_Unet, Res\_Unet, and Res\_Unet++. 


\end{abstract}

\section{Introduction}
The global utilization of Computed Tomography (CT) scans is on the rise. The UNSCEAR 2022 report estimated that globally, from 2009 to 2018, the number of examinations nearly doubled compared to 2006. Chest CT scans accounted for 12.2\% of the total, making them the second most frequently performed type of CT scan, following head CTs \cite{UNSCEAR2022}. With the increase in chest CT examinations, the detection of Anterior Mediastinal Lesions (AMLs) may also rise. Although their frequency is reported to be less than 1\% \cite{yoon2018incidental, henschke2006ct, araki2015anterior}, identifying these lesions on CT is crucial for determining subsequent management \cite{munden2018managing}. With the increase in radiological examinations, radiologist burnout has emerged as a significant issue in the medical field. One potential solution being explored is the use of Artificial Intelligence (AI) \cite{bailey2022understanding}. In the domain of chest radiology, the implementation of comprehensive lung cancer screening programs is anticipated to lead to an increase in the use of Low-dose Chest CT (LDCT) scans \cite{wolf2023screening}. Detecting incidental AMLs on LDCT is important, as these findings may indicate the presence of tumors. This area represents a promising opportunity for radiologists to utilize AI assistance, enhancing their ability to manage workloads and improve diagnostic accuracy.

An automated segmentation model particularly tailored for the AM is a crucial prerequisite for automatically identifying AMLs in the AM, akin to lung cancer screening, which poses a significant hurdle. Retrospectively reviewing chest CT scans over a specific period to investigate the prevalence of AML requires substantial time. Additionally, the lack of commercially available AI models makes utilizing AI for such research difficult. Developing an AI model for AML is particularly challenging because the AM does not have clear boundaries on CT scans, unlike lung parenchyma. This study establishes a hypothetical boundary for the AM and subsequently develops an AM segmentation AI model to assist in investigating the prevalence of AML.

Radiologists can now get assistance utilizing Computer-Aided Detection (CADe) techniques. By highlighting potential organ areas within CT images, CADe methods make screening more efficient and cost-effective \cite{roth2015improving} \cite{castro2020early }. Adhering to segmentation methods is an effective strategy for successfully implementing a CADe system. Segmentation methods are useful in delineating the outlines of lesions and organs, which are essential factors for assessing the malignancy of the lesions \cite{ jung2020differentiating }. These processes emphasize anomalies of CT imaging at the pixel level, which allows them to be distinguished from other procedures. Some methods in \cite{ sandor1991segmentation} and \cite{ye2010automatic} were considered for lesion and organ segmentation, but those methods depend on handwrought features that need several manual procedures.

In recent years, Deep Neural Networks (DNN) have become a professional strategy to solve problems in many areas \cite{ fard2022efficient, torres2020improving, fard2021retinamhsa}, and medical image segmentation is one of them \cite{wang2023anterior, chae2020deep}. The main issues in medical image segmentation for training DNN are lack of dataset, diversity of shapes, domain shift, etc. Using pre-train models, designing lightweight and sophisticated models, and K-fold cross-validation are some methods that are used to solve issues in medical image segmentation.
This work suggests a U shape architecture for segmenting the AM organ from chest CT images to aid radiologists. The proposed model identifies location of AM without any pre-processing or fine-tuning. It employs two distinct attention processes to give a rich feature map. According to the results, the proposed architecture performed better than state-of-the-art (SOTA) segmentation networks involving Res\_Unet\cite{diakogiannis2020resunet}, Res\_Unet++ \cite{jha2019resunet++}, Trans\_Unet \cite{chen2021transunet}, Attention\_Unet \cite{oktay2018attention}, and Unet. The main contributions of this paper are as follows:
\begin{itemize}
  \item This research introduced a U shape structure network for AM segmentation which uses expanded convolution and wide multi-head self-attention in the encoder. This attention effectively allows the model to learn different attention components inside separate subspaces and long range dependencies. 
  \item Specifically, we used Channel Depth-Wise Cross-Correlation Attention (CD-WCC) in the encoder to find similarities between the encoder and decoder to improve organ localization.
  \item The proposed architecture is more lightweight than the SOTA network, with just 6.7 million parameters while it has achieved acceptable and higher performance in all metrics.
  \item To show that our proposed architecture is not limited to our proposed encoder, we used Resnet18 \cite{he2016deep} and the encoder part of the original Unet as the encoder part in the proposed network. These results illustrates the effectiveness of the proposed structure. 
\end{itemize}

The rest of the paper is organized as follows: 
Section~\ref{Related Works} provides a summary of the related research on image segmentation using the Unet framework. In Section~\ref{Proposed Method}, we elaborate each datails of the proposed Unet. Next, Section~\ref{EXPERIMENTS} describes an overview of datasets and the methodologies used to analyze and validate the accuracy of AM segmentation and experimental results. In Section~\ref{Result}, we show the advantage of the proposed network compared to SOTA networks. Finally, the conclusions can be found in Section~\ref{Conclusion}.

\section{Related works}\label{Related Works}
Unet \cite{ronneberger2015u} structure was one of the first DNN networks introduced for segmentation. Many researchers have proposed their segmentation network based on Unet. ViT was proposed for image recognition by Dosovitskiy et al. \cite{dosovitskiy2020image}. Inspired by ViT, Chen et al.\cite{chen2021transunet} introduced Trans\_Unet, which combined transformer and CNN for the encoder part of their proposed Unet.  Trans\_NUnet \cite{yang2022transnunet}, by maintaining the structure of Trans\_Unet in the encoder and using the Convolutional Block Attention Module (CBAM) \cite{woo2018cbam} in the decoder improved performance in Dice score.
In \cite{petit2021u}, fine spatial information was recovered by utilizing Multi-Head Cross-Attention (MHCA) in the decoder. The transformer block is always used in the encoder part to find long-range dependencies, while Petit et al. \cite{petit2021u} implemented the transformer in the decoder to introduce the U-Transformer.

Xu et al. \cite{xu2023levit} utilized LeVit \cite{graham2021levit}, which has four convolution blocks for the encoder part, to propose LeViT Unet. This was the first segmentation work in medical imaging to use transformer architecture to enhance speed. Additionally, they paid attention to potential information by adding attention bias to the attention MLP block.
By replacing traditional convolution layer in Unet by transformer introduced UNETR \cite{hatamizadeh2022unetr}. In their work, transformer extracted features map. Shaker et al. introduced UNETR++ in \cite{shaker2022unetr++} by focusing on channel and spatial attention. Their main idea was an Efficient Paired Attention (EPA) block. While drastically lowering network complexity, UNETR++ outperformed UNETR. Swin UNERT \cite{wang2023swin} used Swin transformer \cite{liu2021swin} for the feature extractor in the encoder part. Based on the Swin transformer's shifting window, their model can learn multi-scale contextual and capture long-range dependencies, outperforming ViT.

Res\_Unet \cite{diakogiannis2020resunet} is a development of the Unet structure, which uses residual connections for the decoder and encoder. In addition, to improve performance, they used an atrous convolution block, multi-tasking inference, and pyramid pooling. Deep residual networks in Resnet architecture and Unet structure were used to design Res\_Unet++ \cite{jha2019resunet++}. The residual block facilitates the transmission of information between layers, enabling the construction of a more profound neural network capable of addressing each encoder’s degradation issue. This enhances the interconnections between channels, thereby decreasing the computational expenditure. The Res\_Unet++ design consists of a stem block, three encoder blocks, an Atrous Spatial Pyramid Pooling (ASPP) module, and three decoder blocks.

Attention methods have been developed to enhance the performance of DNNs by enabling them to concentrate on the most significant aspects of the input. However, based on the input feature map, Multi-Head Self-Attention (MHSA) can be a heavyweight network.
Traditional MHSAs in transformers are the main reason Trans\_Unet is heavyweight. U-Transformer used one MHSA, which has 5 million parameters.
Our proposed network takes a more efficient approach to utilizing MHSAs.
Instead of using one MHSA, we leverage the advantage of parallel MHSAs. By splitting the input feature map into smaller chunks and utilizing parallel MHSAs, we significantly reduce the number of parameter in the proposed network. 

\section{Proposed Network Architecture} \label{Proposed Method}
\begin{figure*}[bt]
    \centering
    \includegraphics[width=15cm]{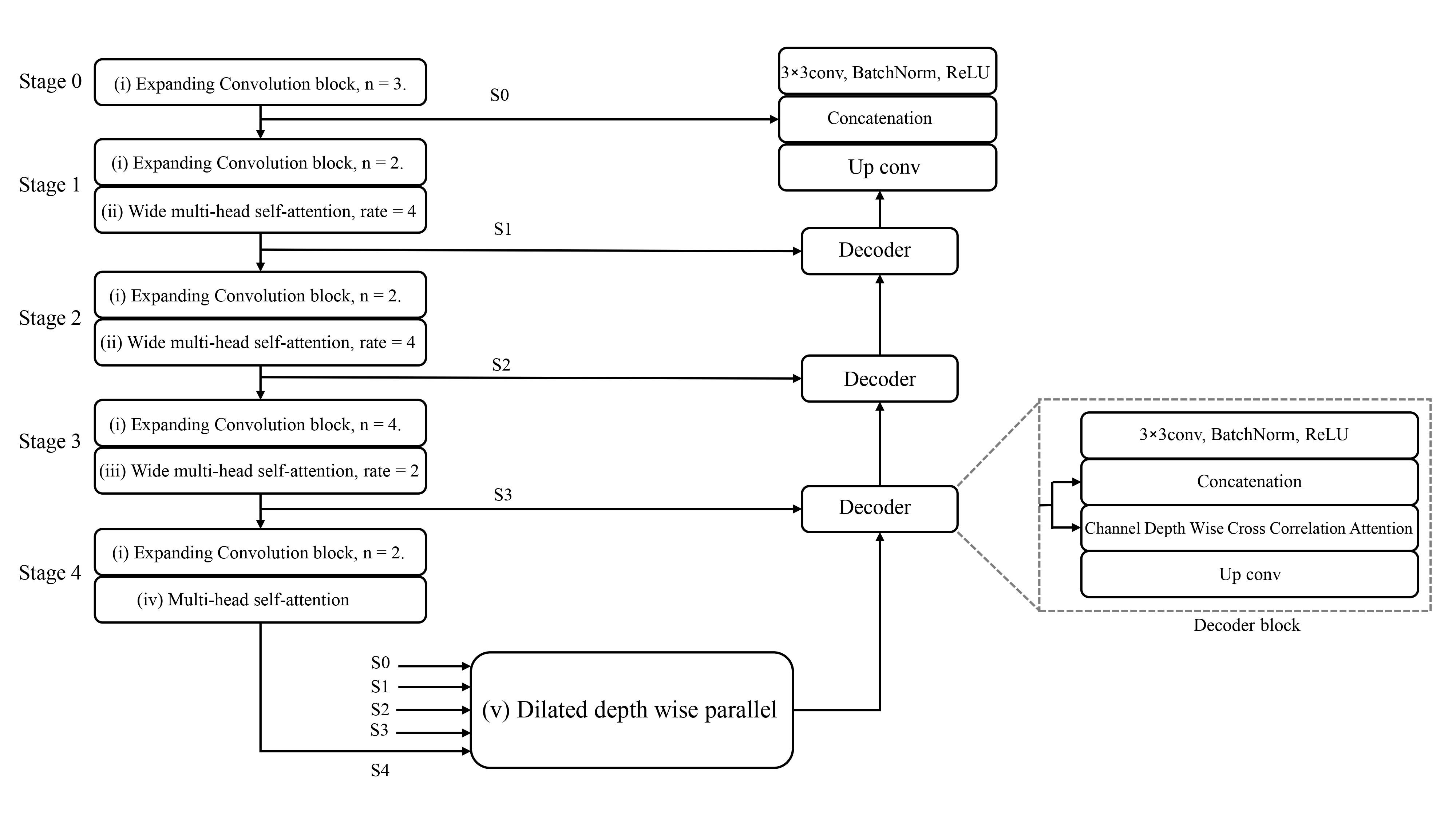}
    
    \caption{Proposed network for anterior mediastinum segmentation. (i): Expanding convolution block introduced in Fig.~\ref{fig2}, n refers to amount of 1×1 based on input feature map. (ii): Fig. ~\ref{fig3} (a) is related to four parallel MHSAs used for the proposed network. (iii): W-MHSA with rate 2 is utilized two MHSA in the parallel way shown in Fig. ~\ref{fig3} (b). (iv): Traditional MHSA shown in Fig. ~\ref{fig3} (c). (v): Delited depth-wise parallel shown in Fig. ~\ref{fig5}.}
    \label{figure1}
\end{figure*}

The proposed architecture shown in Fig.~\ref{figure1} was designed in U shape structure which has three components: encoder, decoder, and parallel path connection between encoder and decoder. The encoder extracts feature maps and decreases input dimensions from 3×224×224 to 512×7×7. The decoder improves feature maps' dimensions relative to the input dimension size. Both the encoder and decoder are designed into five sequential stages. Moreover, we employ two highly effective attention methods to significantly bolster the restoration of organ segmentation in each decoder and encoder block, instilling confidence in the model's capabilities. 
After improving height and width in decoder, channel attention applies to channels dimension. Instead of using average and maximum in height and width for channel attention, we utilize depth-wise cross correlation \cite{li2019siamrpn++} in order to find similarity feature maps between each stage in encoder and decoder. In the next step, the effective of similarity multiply to feature maps. Parallel path connection is designed into five dilated convolutions with different filter size and dilation rate. Dilated convolutions provide advantages such as enhanced receptive field size and retrieving features at several scales.

\subsection{Expanding Convolution block}
\begin{figure}[bt]
    \centering
    \includegraphics[width=8cm]{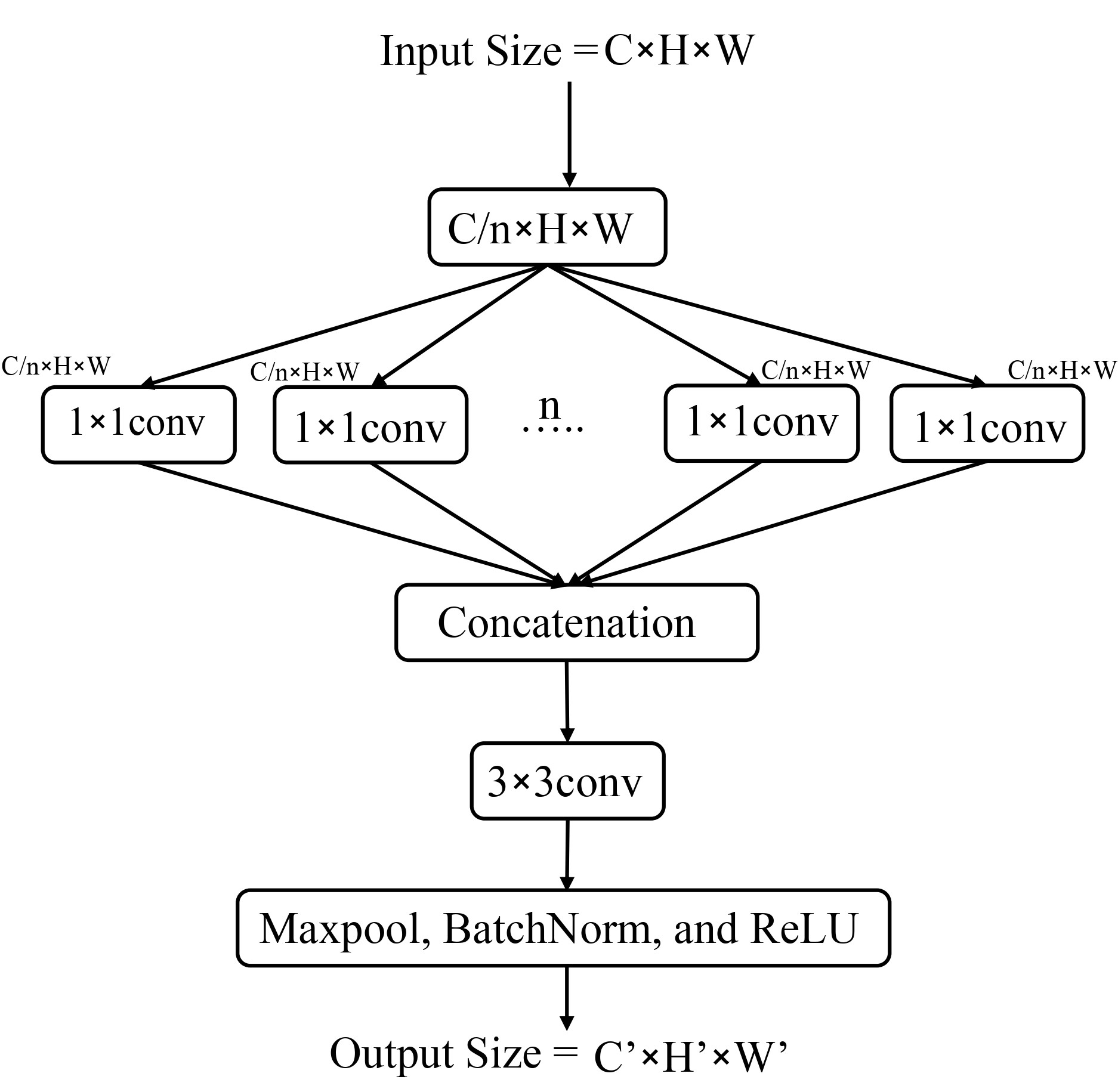}
    
    \caption{Expanding convolution block (shown in Fig. ~\ref{figure1}). The feature map channel is divided into small chunks according to the number of n for each stage: stage 0 = 3, stage 1 = 2, stage 2 = 2, stage 3 = 4, stage 4 = 2.}
    \label{fig2}
\end{figure}

Going wider into the feature extraction section, we expand the convolutional layer and build the proposed feature extraction in five stages which is illustrated in Fig.~\ref{fig2}. In addition, 1×1 convolutions are employed as opposed to 3×3 convolutions as main part of extract features. The input channel for every step is separated into smaller chunks, and then a separate convolution filter is applied independently for each smaller channel. In order to be ready for the next stage, the specific features from all channels are concatenated in channel dimension and then pass through one 3×3 convolutional layer. The input image size is 3×224×224. 
\begin{itemize}
    \item In stage 0, according to the input channel, three different 1×1 convolution layers are applied independently. After concatenation, one 3×3 convolution layer improves channel to 64 while height and width decrease to 112.
    \item In stage 1, 64 channels are divided into 2. Two 32×112×112 passes through the 1×1 convolution layers simultaneously and are concatenated in the channel dimension to return to 64 channels then resulting in a feature extractor size will be 128×56×56.
    \item The exact structure, like stage 1, is considered for stage 2. However, the output size's channel, height, and width dimensions will be 128, 28, and 28, respectively.
    \item Stage 3 is designed to be deeper and broader than other stages to capture more complicated features and patterns as the input data progresses through the network. Although it maintains the same structure as previous stages, it involves four 1×1 convolutions simultaneously, concatenation, and one additional 3×3 convolution. Finally, the feature extraction from stage 3 results in a size of 256×14×14.
    \item The strategy for designing stage 4 is the same as the previous stage. However, the input channel size is 256, divided by 2. Therefore, two 64×14×14 enter to two 1×1 convolutions independently. The output size will be 512×7×7 after concatenation, passing from one 3×3 convolution.
\end{itemize}

\subsection{Wide Multi-Head Self-Attention (W-MHSA)}
\begin{figure*}[bt]
     \centering
     \begin{subfigure}[b]{0.35\textwidth}
         \centering
         \includegraphics[width=\textwidth]{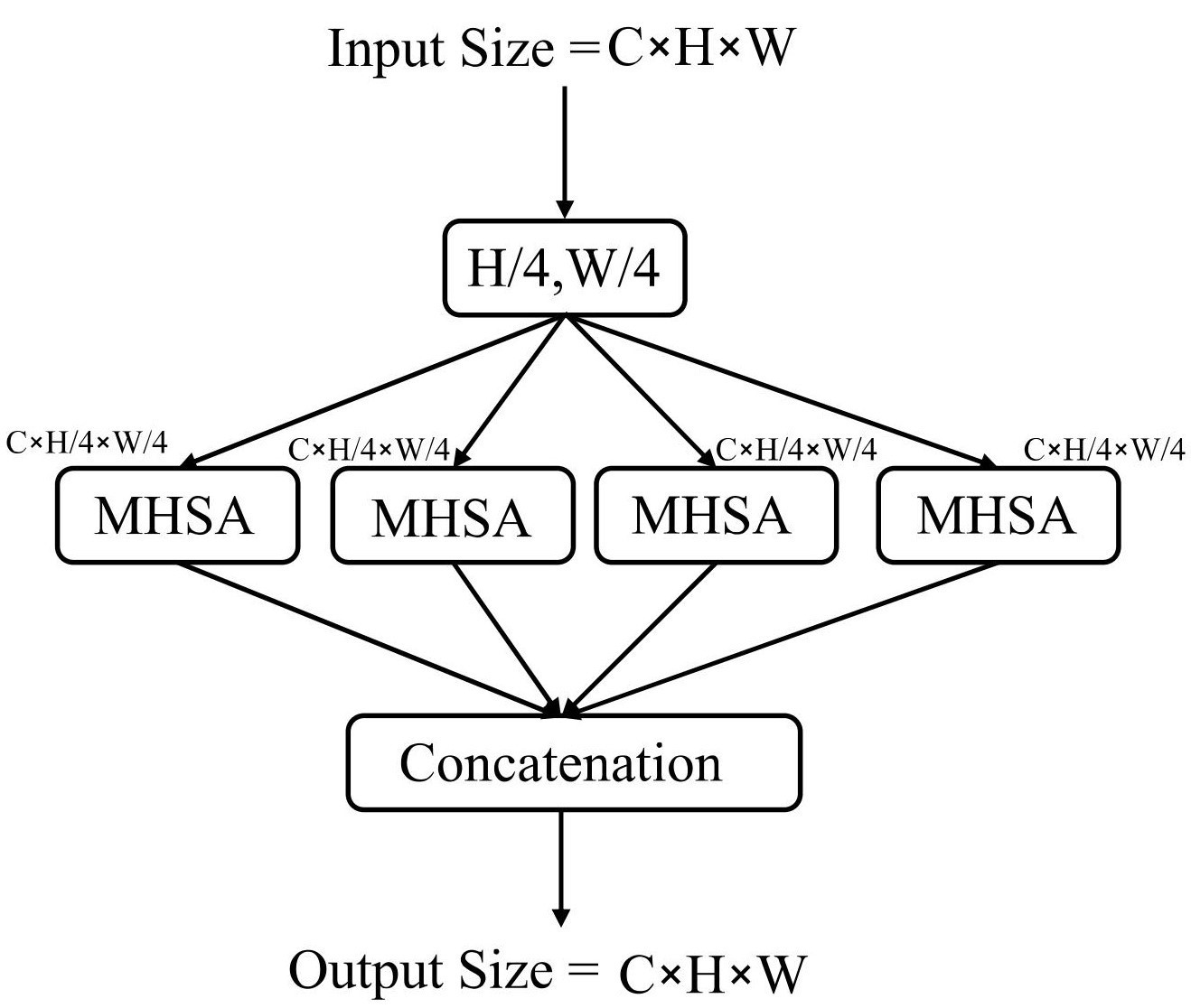}
         \caption{W-MHSA, four parallel ways}
     \end{subfigure}\hspace{0.1cm}%
     \hfill
     \begin{subfigure}[b]{0.25\textwidth}
         \centering
         
         \includegraphics[width=\textwidth]{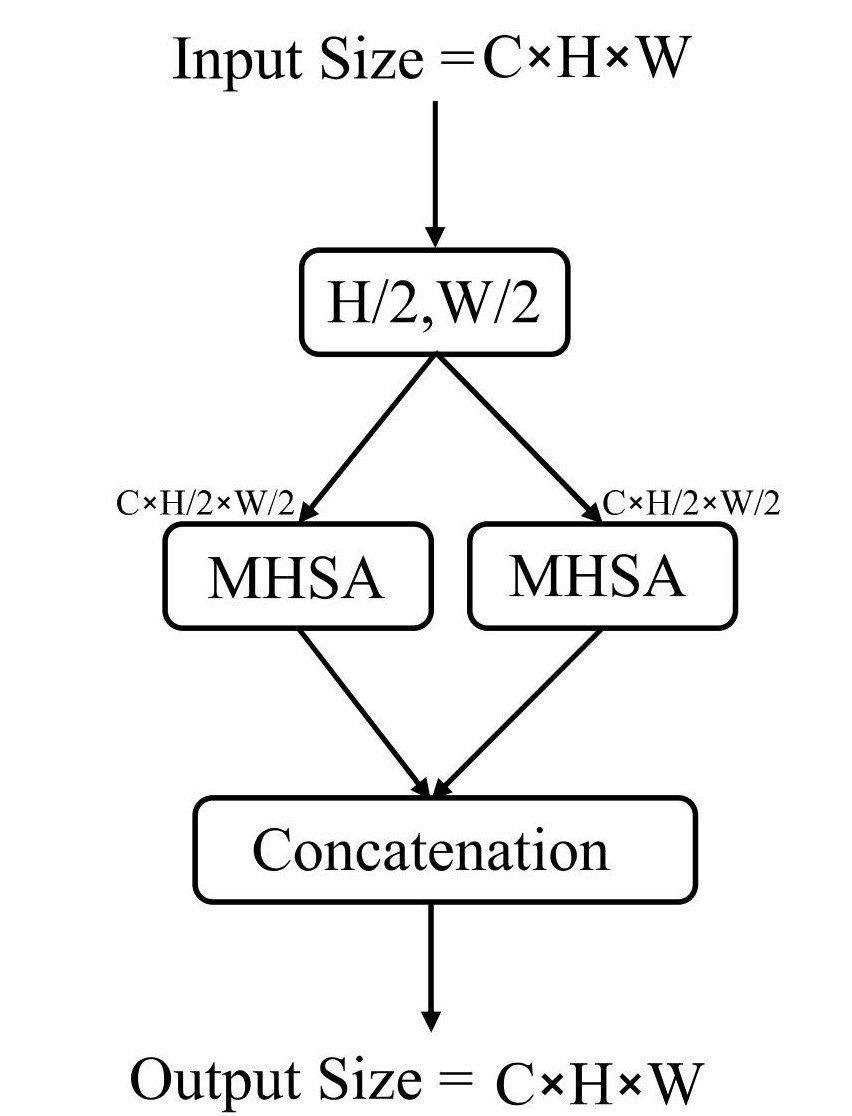}
         \caption{W-MHSA, two parallel ways}
     \end{subfigure}\hspace{0.3cm}%
     \hfill
     \begin{subfigure}[b]{0.15\textwidth}
         \centering
         \includegraphics[width=\textwidth]{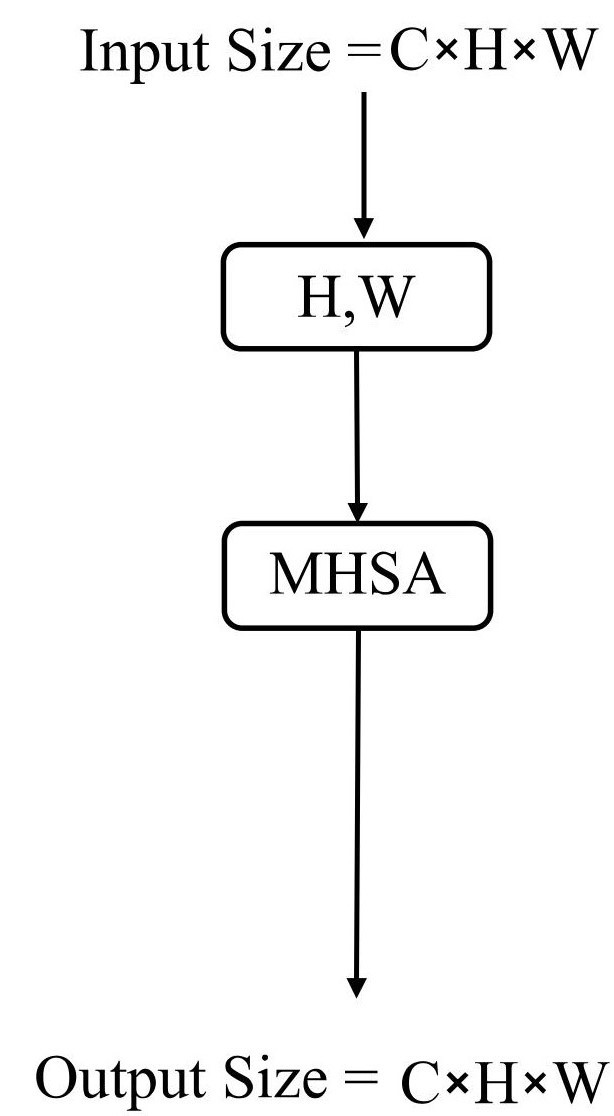}
         \caption{MHSA}
     \end{subfigure}\hspace{0.1cm}%
     \hfill
        \caption{Attention with different parallel paths. (a): It is designed with four MHSA (rate of width = 4) (shown in (ii) in Fig. ~\ref{figure1}), (b): It is designed with two MHSA (rate of width = 2) (shown in (iii) in Fig. ~\ref{figure1}), and (c): It is designed with one MHSA (shown in (iv) in Fig. ~\ref{figure1}).}
        \label{fig3}
\end{figure*}

The attention mechanism can be heavy weight according to the input dimension because due to fully connected layer. In this study, height and width of feature maps is divided into small chunks to overcome the heavy weight of self-attention. Three different MHSA are utilized in encoder of the proposed network. Four parallel MHSAs shown in (Fig.~\ref{fig3}(a)), two parallel MHSAs shown in (Fig.~\ref{fig3}(b)), and one MHSA shown in (Fig.~\ref{fig3}(c)) are used in stages 1 and 2, stage 3, and stage 4, respectively. Parallel MHSAs are named Wide-MHSA (W-MHSA) in this study. 
\begin{itemize}
    \item After passing the input from stage 1, the output will be 64×56×56. First, W-MHSA (rate of width = 4) is applied after stage 1. W-MHSA (rate of width = 4) is designed in four branches. Therefore, the input for W-MHSA (rate of width = 4) is divided into 4 in height and width. First, four 64×14×14 enter to four different MHSA. Second, four outputs from MHSAs are concatenated in height and width dimensions. These outputs from concatenations are again concatenated by height and weight, and then two concatenation phases multiply together. Finally, it passes from one convolution layer.
    \item Furthermore, there is a comparable procedure for stage 2. The input 128×28×28 is split by 4. Subsequently, four 128×7×7 enter four separate MHSAs using a methodology similar to stage 1.
    \item After stage 3 is completed, the input size is 256×14×14. The steps are the same as in the previous phases, with the exception that the width and height are split in two. 2 × (256×7×7) is passed from each of the two parallel MHSAs employed (W-MHSA(rate of width = 2)). Height and width are concatenated from two MHSA outputs in height and width dimension. Then, these two concatenations are concatenated again regarding height and width dimensions. Finally, the last two concatenations are multiplied and passed to a convolution layer.
    \item In addition, one MHSA is used for after stage 4 which the input dimension is 512×7×7. 
    We provide more details of W-MHSA in Fig.~\ref{fig3}, in which (a) divides input size to 4, (b) divides to 2, and (c) is the original MHSA.
\end{itemize}
    

\subsection{Channel Depth Wise Cross Correlation Attention (CDWCC)}
\begin{figure*}[bt]
    \centering
    \includegraphics[width=12cm]{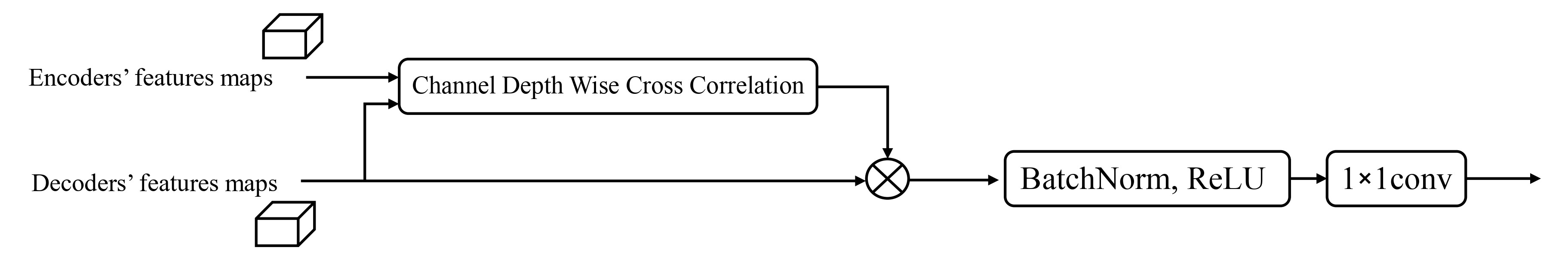}
    
    \caption{Channel Depth Wise Cross Correlation Attention (shown in decoder part of Fig. ~\ref{figure1})}
    \label{fig4}
\end{figure*}

The cross-correlation module is a computational process that convolves two sets of feature maps. It is typically employed for tracking purposes in techniques like SiamRPN++ \cite{li2019siamrpn++}. The mechanism of working in this module is similar to that of two distinctive feature maps. This study uses cross-correlation to find similarities between the encoder and decoder. The correlation procedure is performed channel by channel on the two different feature maps with an equal number of channels, height, and width. According to Fig. ~\ref{fig4}, the decoders’ features map is input for CDWCC, while the encoders’ features map is a kernel for performing the correlation operation channel by channel to find similarity. Finally, the CDWCC feature map is multiplied element-wise with the decoder feature map.

\subsection{Dilated depth wise parallel path connection (DDWPP)}
\begin{figure*}[bt]
    \centering
    \includegraphics[width=13cm]{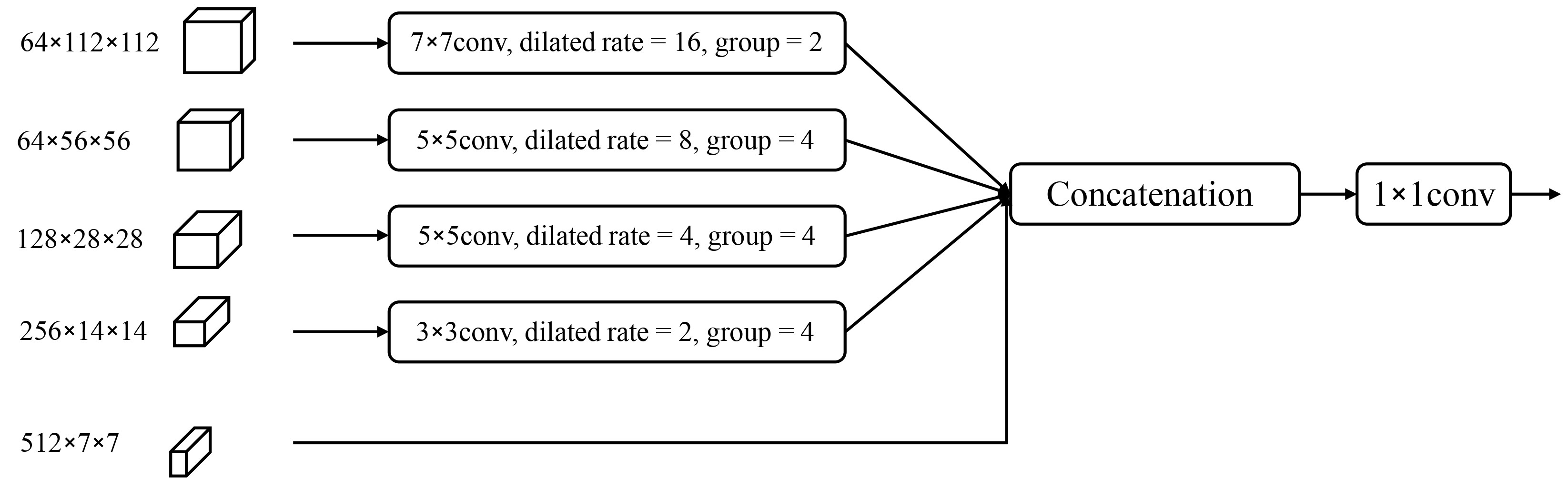}
    
    \caption{Dilated depth wise parallel path connection (shown in (v) in Fig. ~\ref{figure1}). The input channels are convolved separately with their own sets of filters, each based on a distinct number of groups.}
    \label{fig5}
\end{figure*}

In order to have high field of view, DDWPP is designed. As shown in Fig. \ref{fig5}, it has five parallel dilated convolutions with different kernel size and dilated rate which capture large receptive fields without increasing the number of parameters. In the process of segmentation, down sampling feature maps can potentially discard crucial components. Therefore, preserving information from the initial stages of the encoder is imperative for the subsequent up sampling phase. Having feature maps of each stage effectively helps our proposed network avoid missing information. In addition, primary stages have low level information while the latest stages have valuable one. As seen in Fig. \ref{fig5}, the dilated convolution technique expands the receptive field while maintaining the spatial precision of the feature maps. We have used four parallel dilated convolutions in the DDWPP block to increase the detection sensitivity for small AMs. The four dilated convolutions follow dilated rates of 16, 8, 4, and 2, and filter sizes of 7×7, 5×5, 5×5, and 3×3, respectively.

\section{Materials and Experiments} \label{EXPERIMENTS}
This study was approved by the Institutional Review Board of Chonnam National University Hwasun Hospital (CNUHH-2022-241). Informed consent was waived due to its retrospective design.
\subsection{Dataset}
This study was conducted with 200 patients who underwent chest CT as part of a health check-up at CNUHH between September 2020 and December 2021, and who exhibited no AMLs in the anterior mediastinum. A thoracic radiologist with 5-years of experience in chest radiology (W.G.J.) visually reviewed all chest CT scans and selected those without AMLs. The mean age of the patients was 55.9 years (SD, 10.0 years), and the cohort comprised 120 males (60\%) and 80 females (40\%).
We defined anterior mediastinum as following criteria: (i) superior border: thoracic inlet, (ii) Inferior border: base of pulmonary trunk, (iii) anterior border: posterior margin of sternum, (iv) posterior border: anterior aspect of pericardium, and (v) lateral border: parietal mediastinal pleura \cite{yoon2018incidental}. Detailed CT parameters are described in the Table ~\ref{CT_table}.

\begin{table*}[h]
\centering
\caption{Detailed CT parameters}
\resizebox{\textwidth}{!}{%
  \begin{tabular}{|c|c|}
    \hline
    \textbf{CT machine (Vendor)} &  \textbf{Parameters} \\
    \hline
    & Section thickness: 2.5--5.0 mm \\  
    & Rotation time: 0.5 s \\            
    & Peak kilovoltage: 120 kVp \\
    Revolution HD (GE Healthcare, Waukesha, WI, USA) & Tube current: 60--220 mAs, with an automatic \\
    & exposure control \\
    & Kernel: standard \\
    & Reconstruction algorithm: iterative reconstruction \\
    \hline
    & Section thickness: 3.0--4.0 mm \\  
    & Rotation time: 0.5 s \\            
    & Peak kilovoltage: 100 kVp, 120 kVp \\
    Somatom Definition Flash (Siemens Healthineers, Erlangen, Germany) & Tube current: 60--220 mAs, with an automatic \\
    & exposure control \\
    & Kernel: B31f \\
    & Reconstruction algorithm: iterative reconstruction \\
    \hline
  \end{tabular}
}
\end{table*}

\begin{figure*}[bt]
\centering
\begin{minipage}{0.7\columnwidth}
  \centering
  \includegraphics[width=\linewidth]{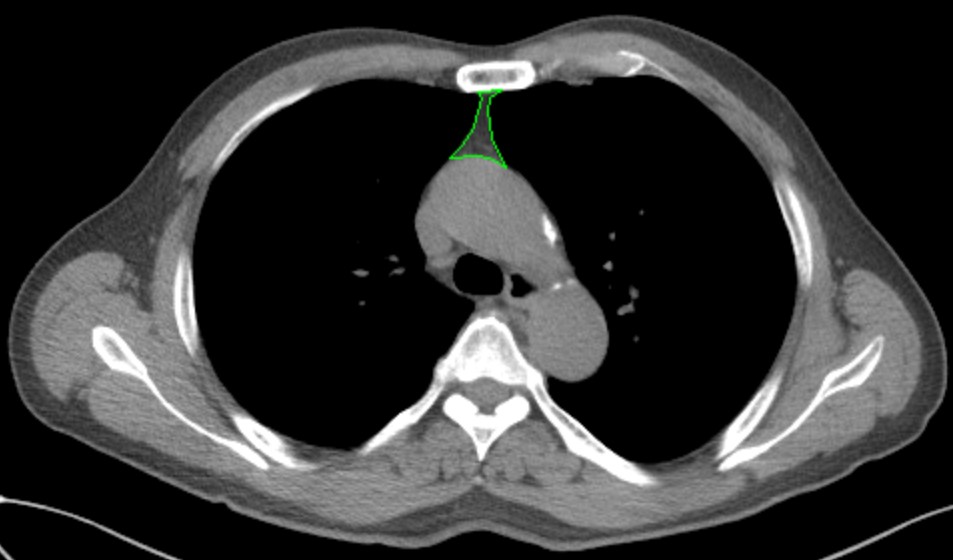}
\end{minipage}%
\begin{minipage}{0.7\columnwidth}
  \centering
  \includegraphics[width=\linewidth]{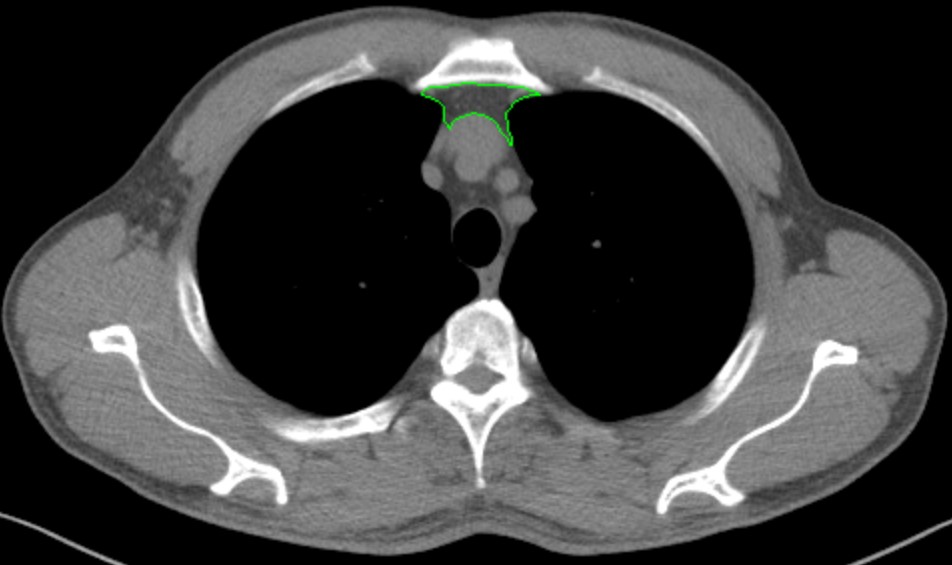}
\end{minipage}

\vspace{0.05cm}

\caption{Two distinct anterior mediastinal instances from one patient are shown inside the green border to illustrate diverse shape of AM.}
\label{fig6}
\end{figure*}

2776 slices from 200 patients were considered for training the Deep Learning (DL) model. Initially, we transformed Digital Imaging and Communications in Medicine (DICOM) into PNG format, a technological standard for digitally storing and transmitting medical images. We considered the traditional mediastinal window settings, which included a window level of 30 and a window width of 520 Houndsfield Units (HU).  Finally, the generated PNG images had resolutions of 512×512, and a radiologist manually drew masks for AM using the labelme \cite{labelmeai} annotation tool. 
As shown in Fig. \ref{fig6}, an AM present in two slice images has distinct visual depictions with resolutions of 512×512, HU of 520, and window level of 30.

\subsection{Evaluation metrics}
Our study utilizes Dice Similarity Coefficient (DSC), mean Intersection over Union (IoU), sensitivity, and Accuracy (Acc), to comprehensively examine and evaluate the effectiveness of our SOTA research technique.
The equations representing DCS, IoU, sensitivity, and Acc are expressed as equations (1), (2), (3), and (4) accordingly. The metrics are computed based on the values of True Positive (TP), False Positive (FP), True Negative (TN), and False Negative (FN). TP and TN represent the accuracy of positive and negative predictions, respectively. FP and FN indicate the ratio of positive predictions that were wrongly forecasted and negative predictions that were incorrectly predicted, respectively.

\begin{equation}
Dice=\frac{2×TP}{2×TP+FP+FN}
\end{equation} \label{DSC}

\begin{equation}
IoU=\frac{TP}{TP+FP+FN}
\end{equation} \label{iou}

\begin{equation}
Sensitivity=\frac{TP}{TP+FP}
\end{equation}

\begin{equation}
Accuracy=\frac{TN+TP}{TN+TP+FN+FP}
\end{equation}
\subsection{Implementation details}
For both the training and validation phases, we used 4-fold cross-validation, with each model being trained using 100 epochs per fold. The loss function used is dice loss \cite{milletari2016v}, in which loss is optimized by utilizing Adam with an initial learning rate of 0.0003. The learning rate was decreased by 10 times in epochs 25 and 180. In addition, 32 batch sizes were considered.
Pytorch framework was used to implement the proposed and other networks on A40 GPU.


\begin{figure*}[bt]
     \centering
     \begin{subfigure}[b]{0.12\textwidth}
         \centering
         \includegraphics[width=\textwidth]{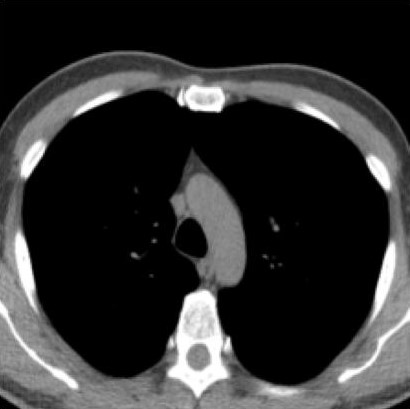}
         
     \end{subfigure}\hspace{0.05cm}%
     \hfill
     \begin{subfigure}[b]{0.12\textwidth}
         \centering
         \caption*{Ground truth}
         \includegraphics[width=\textwidth]{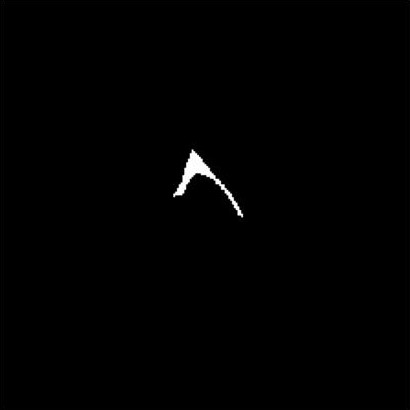}
         
     \end{subfigure}\hspace{0.05cm}%
     \hfill
     \begin{subfigure}[b]{0.12\textwidth}
         \centering
         \caption*{Unet}
         \includegraphics[width=\textwidth]{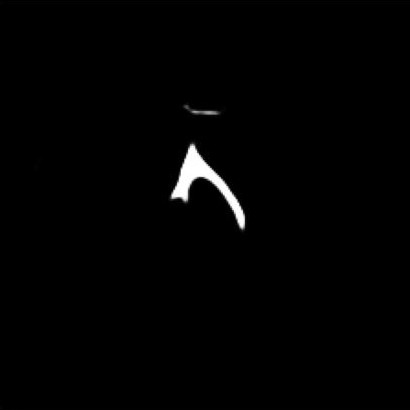}
         
     \end{subfigure}\hspace{0.05cm}%
     \hfill
     \begin{subfigure}[b]{0.12\textwidth}
         \centering
         \caption*{Res\_Unet}
         \includegraphics[width=\textwidth]{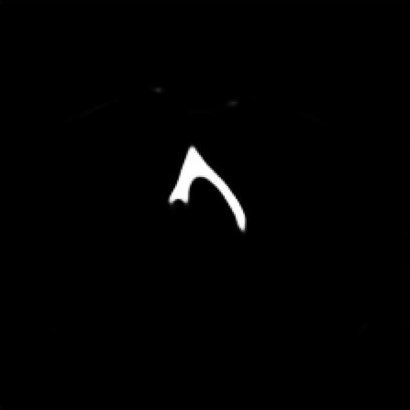}
    \end{subfigure}\hspace{0.05cm}%
     \hfill
     \begin{subfigure}[b]{0.12\textwidth}
         \centering
         \caption*{Res\_Unet++}
         \includegraphics[width=\textwidth]{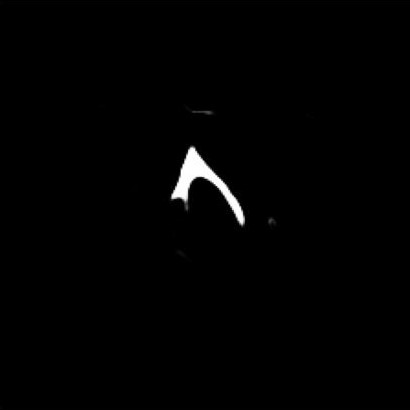}

     \end{subfigure}\hspace{0.05cm}%
     \hfill
     \begin{subfigure}[b]{0.12\textwidth}
         \centering
         \caption*{Attention\_Unet}
         \includegraphics[width=\textwidth]{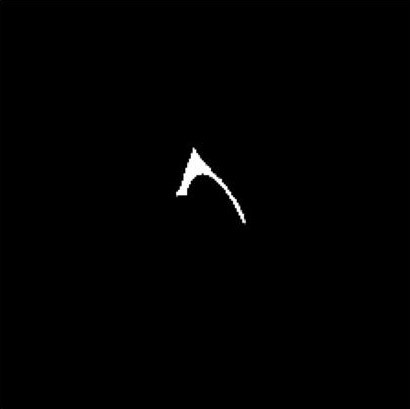}
         
     \end{subfigure}\hspace{0.05cm}%
     \hfill
     \begin{subfigure}[b]{0.12\textwidth}
         \centering
         \caption*{Trans\_Unet}
         \includegraphics[width=\textwidth]{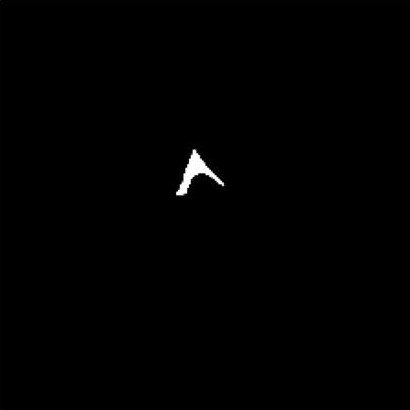}
         
     \end{subfigure}\hspace{0.05cm}%
     \hfill
     \begin{subfigure}[b]{0.12\textwidth}
         \centering
         \caption*{Proposed\_Unet}
         \includegraphics[width=\textwidth]{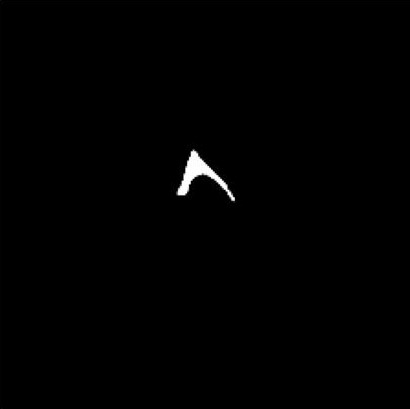}
         
     \end{subfigure}
     
     \medskip 
    \begin{subfigure}[b]{0.12\textwidth}
         \centering
         \includegraphics[width=\textwidth]{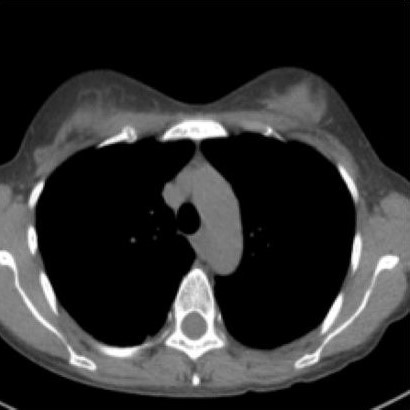}
         
     \end{subfigure}\hspace{0.05cm}%
     \hfill
     \begin{subfigure}[b]{0.12\textwidth}
         \centering
         
         \includegraphics[width=\textwidth]{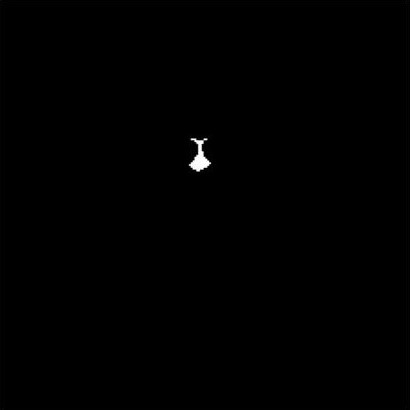}
         
     \end{subfigure}\hspace{0.05cm}%
     \hfill
     \begin{subfigure}[b]{0.12\textwidth}
         \centering
         
         \includegraphics[width=\textwidth]{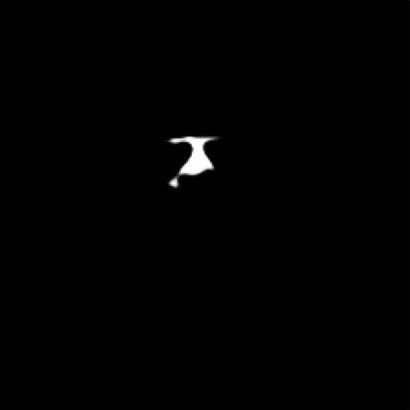}
         
     \end{subfigure}\hspace{0.05cm}%
     \hfill
     \begin{subfigure}[b]{0.12\textwidth}
         \centering
         
         \includegraphics[width=\textwidth]{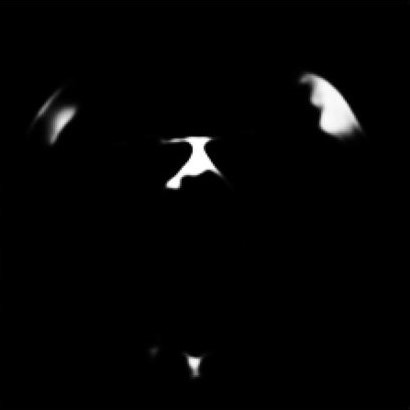}
    \end{subfigure}\hspace{0.05cm}%
     \hfill
     \begin{subfigure}[b]{0.12\textwidth}
         \centering
         
         \includegraphics[width=\textwidth]{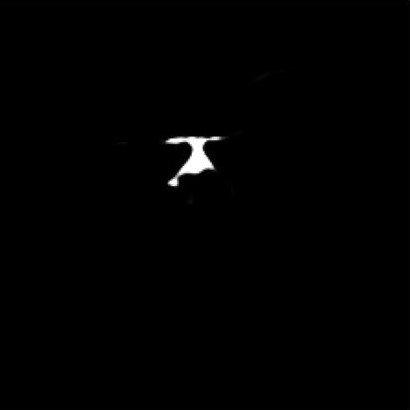}

     \end{subfigure}\hspace{0.05cm}%
     \hfill
     \begin{subfigure}[b]{0.12\textwidth}
         \centering
         
         \includegraphics[width=\textwidth]{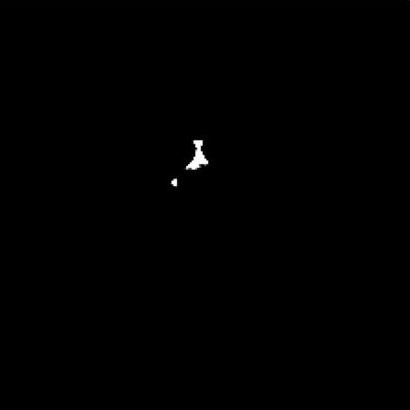}
         
     \end{subfigure}\hspace{0.05cm}%
     \hfill
     \begin{subfigure}[b]{0.12\textwidth}
         \centering
         
         \includegraphics[width=\textwidth]{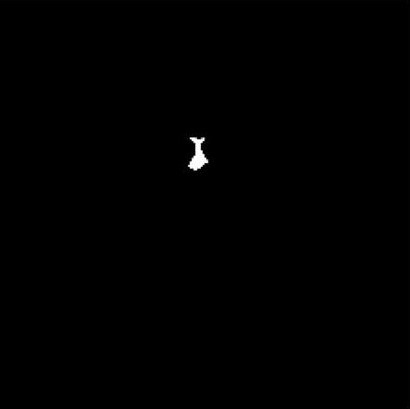}
         
     \end{subfigure}\hspace{0.05cm}%
     \hfill
     \begin{subfigure}[b]{0.12\textwidth}
         \centering
         
         \includegraphics[width=\textwidth]{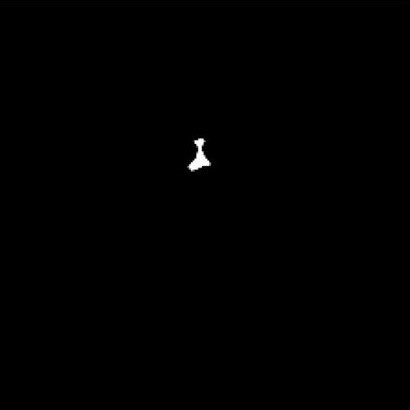}
    
     \end{subfigure}
     
     \medskip 
    \begin{subfigure}[b]{0.12\textwidth}
         \centering
         \includegraphics[width=\textwidth]{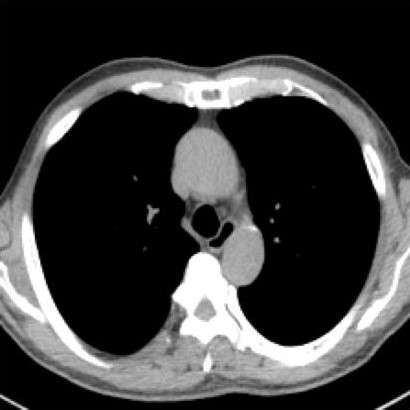}
         
     \end{subfigure}\hspace{0.05cm}%
     \hfill
     \begin{subfigure}[b]{0.12\textwidth}
         \centering
         
         \includegraphics[width=\textwidth]{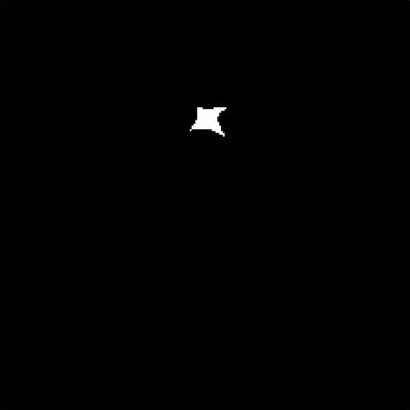}
         
     \end{subfigure}\hspace{0.05cm}%
     \hfill
     \begin{subfigure}[b]{0.12\textwidth}
         \centering
         
         \includegraphics[width=\textwidth]{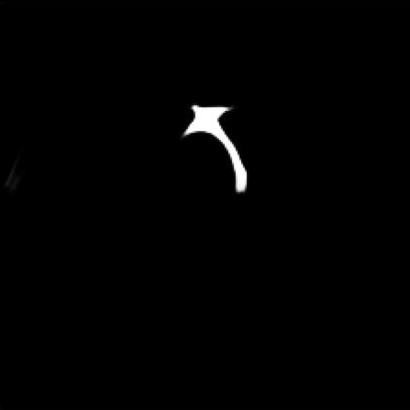}
         
     \end{subfigure}\hspace{0.05cm}%
     \hfill
     \begin{subfigure}[b]{0.12\textwidth}
         \centering
         
         \includegraphics[width=\textwidth]{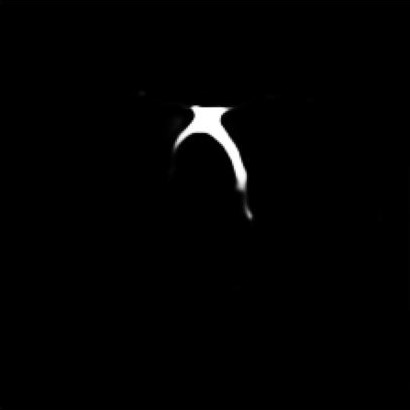}
    \end{subfigure}\hspace{0.05cm}%
     \hfill
     \begin{subfigure}[b]{0.12\textwidth}
         \centering
         
         \includegraphics[width=\textwidth]{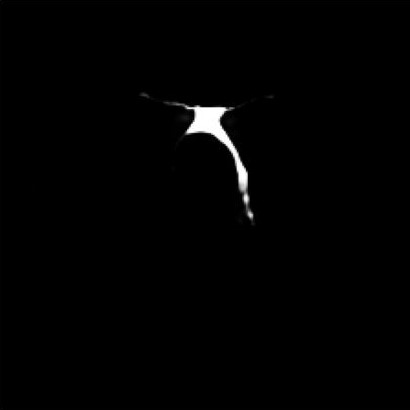}

     \end{subfigure}\hspace{0.05cm}%
     \hfill
     \begin{subfigure}[b]{0.12\textwidth}
         \centering
         
         \includegraphics[width=\textwidth]{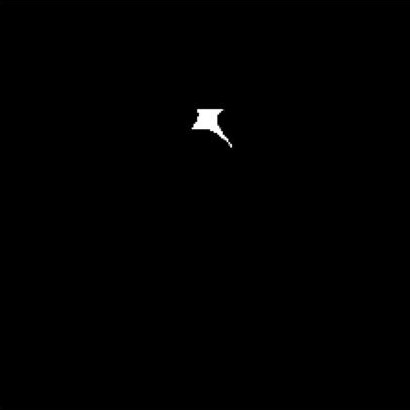}
         
     \end{subfigure}\hspace{0.05cm}%
     \hfill
     \begin{subfigure}[b]{0.12\textwidth}
         \centering
         
         \includegraphics[width=\textwidth]{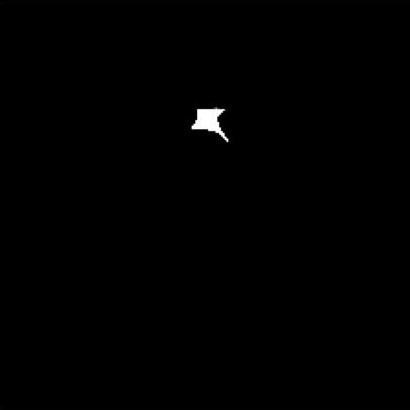}
         
     \end{subfigure}\hspace{0.05cm}%
     \hfill
     \begin{subfigure}[b]{0.12\textwidth}
         \centering
         
         \includegraphics[width=\textwidth]{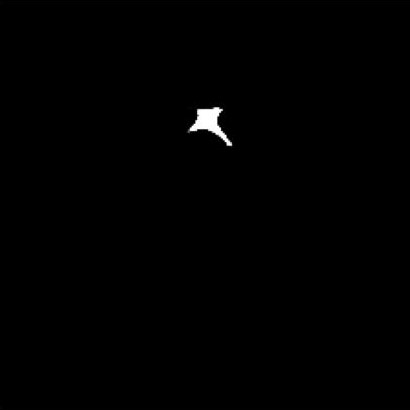}
         
     \end{subfigure}
     
     \medskip 
    \begin{subfigure}[b]{0.12\textwidth}
         \centering
         \includegraphics[width=\textwidth]{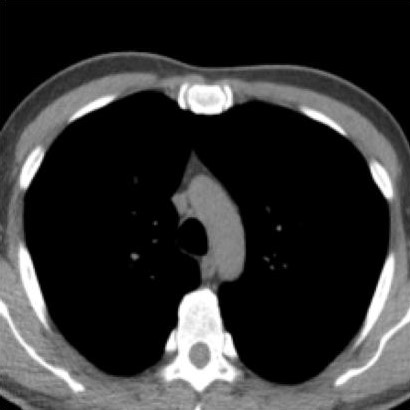}
         
     \end{subfigure}\hspace{0.05cm}%
     \hfill
     \begin{subfigure}[b]{0.12\textwidth}
         \centering
         
         \includegraphics[width=\textwidth]{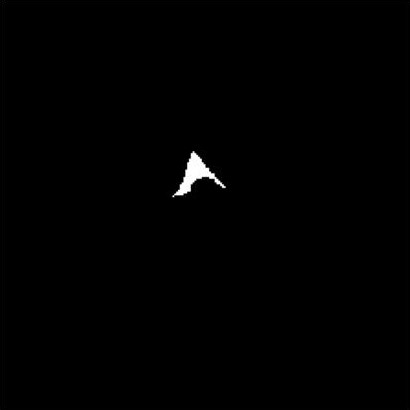}
         
     \end{subfigure}\hspace{0.05cm}%
     \hfill
     \begin{subfigure}[b]{0.12\textwidth}
         \centering
         
         \includegraphics[width=\textwidth]{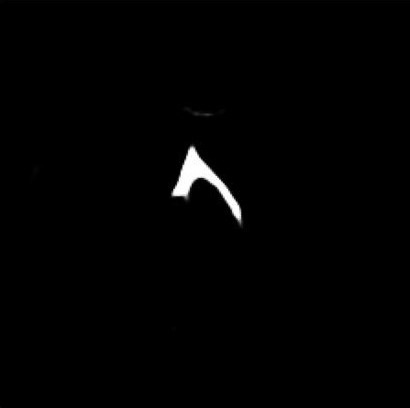}
         
     \end{subfigure}\hspace{0.05cm}%
     \hfill
     \begin{subfigure}[b]{0.12\textwidth}
         \centering
         
         \includegraphics[width=\textwidth]{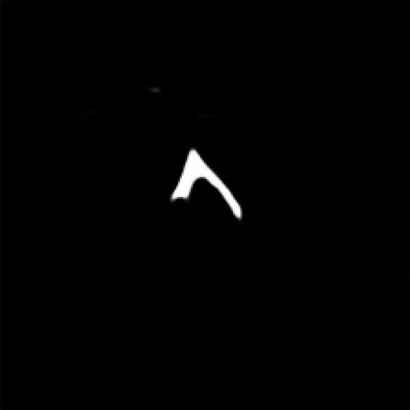}
    \end{subfigure}\hspace{0.05cm}%
     \hfill
     \begin{subfigure}[b]{0.12\textwidth}
         \centering
         
         \includegraphics[width=\textwidth]{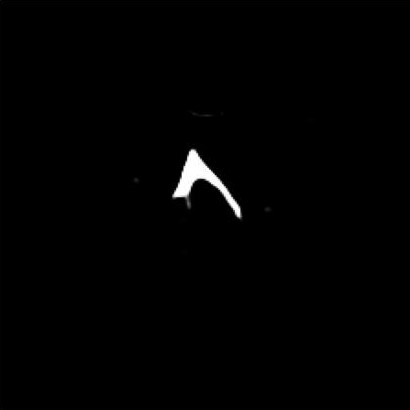}

     \end{subfigure}\hspace{0.05cm}%
     \hfill
     \begin{subfigure}[b]{0.12\textwidth}
         \centering
         
         \includegraphics[width=\textwidth]{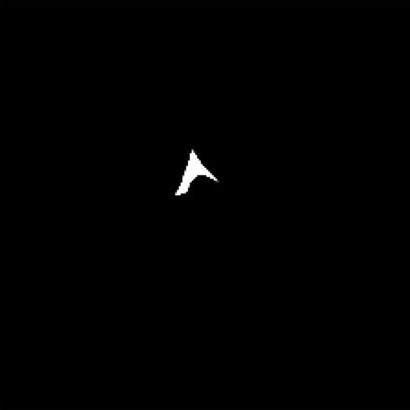}
         
     \end{subfigure}\hspace{0.05cm}%
     \hfill
     \begin{subfigure}[b]{0.12\textwidth}
         \centering
         
         \includegraphics[width=\textwidth]{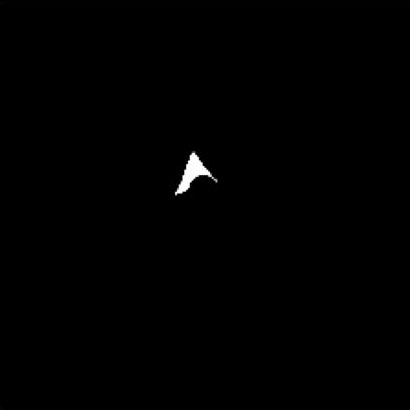}
         
     \end{subfigure}\hspace{0.05cm}%
     \hfill
     \begin{subfigure}[b]{0.12\textwidth}
         \centering
         
         \includegraphics[width=\textwidth]{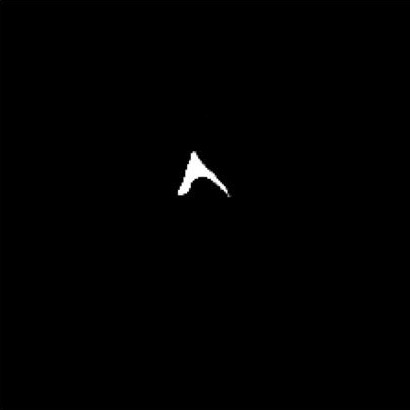}
         
     \end{subfigure}
          
     \medskip 
    \begin{subfigure}[b]{0.12\textwidth}
         \centering
         \includegraphics[width=\textwidth]{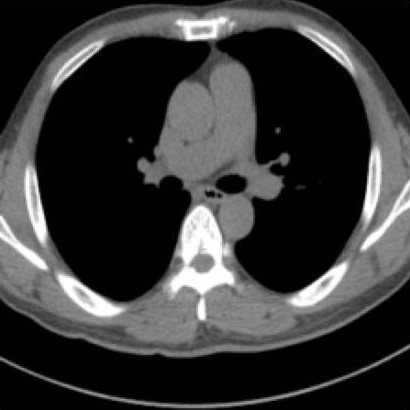}
         
     \end{subfigure}\hspace{0.05cm}%
     \hfill
     \begin{subfigure}[b]{0.12\textwidth}
         \centering
         
         \includegraphics[width=\textwidth]{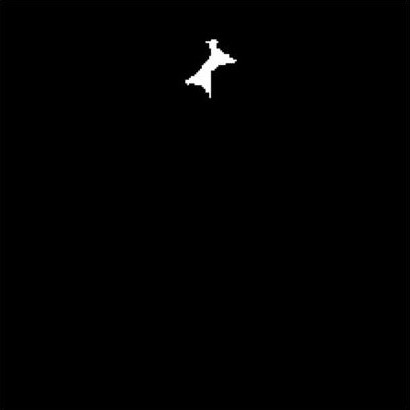}
         
     \end{subfigure}\hspace{0.05cm}%
     \hfill
     \begin{subfigure}[b]{0.12\textwidth}
         \centering
         
         \includegraphics[width=\textwidth]{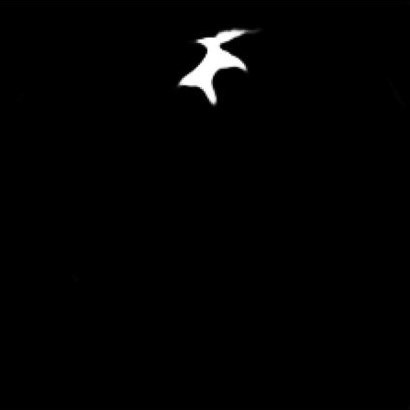}
         
     \end{subfigure}\hspace{0.05cm}%
     \hfill
     \begin{subfigure}[b]{0.12\textwidth}
         \centering
         
         \includegraphics[width=\textwidth]{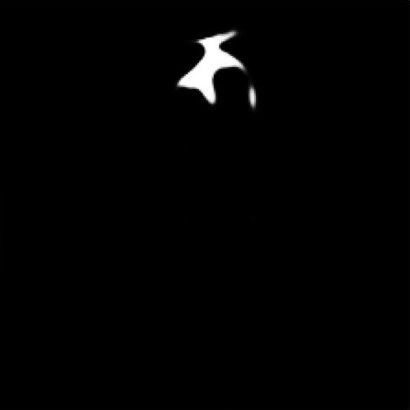}
    \end{subfigure}\hspace{0.05cm}%
     \hfill
     \begin{subfigure}[b]{0.12\textwidth}
         \centering
         
         \includegraphics[width=\textwidth]{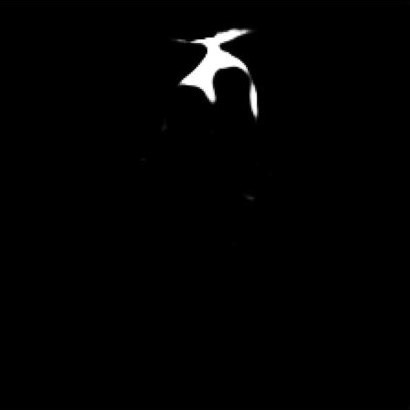}

     \end{subfigure}\hspace{0.05cm}%
     \hfill
     \begin{subfigure}[b]{0.12\textwidth}
         \centering
         
         \includegraphics[width=\textwidth]{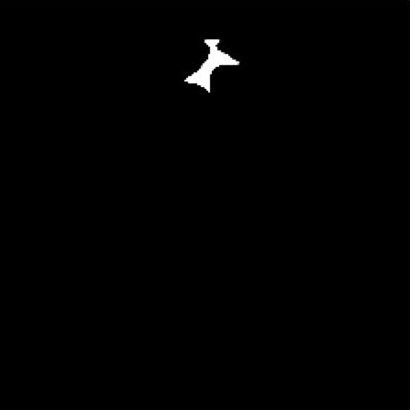}
         
     \end{subfigure}\hspace{0.05cm}%
     \hfill
     \begin{subfigure}[b]{0.12\textwidth}
         \centering
         
         \includegraphics[width=\textwidth]{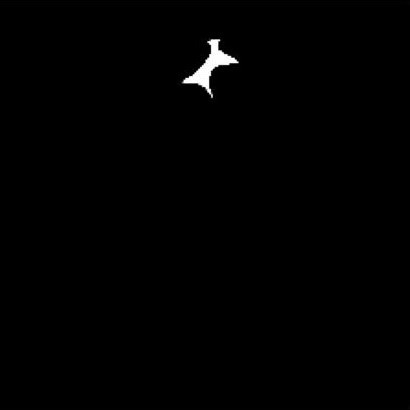}
         
     \end{subfigure}\hspace{0.05cm}%
     \hfill
     \begin{subfigure}[b]{0.12\textwidth}
         \centering
         
         \includegraphics[width=\textwidth]{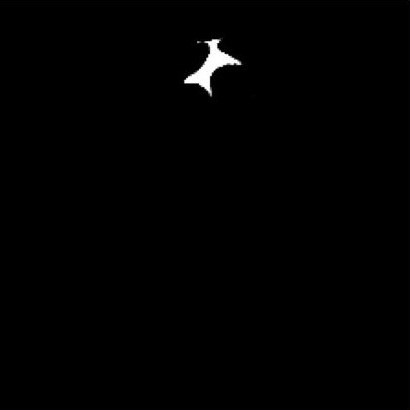}
         
     \end{subfigure}
          
     \medskip 
    \begin{subfigure}[b]{0.12\textwidth}
         \centering
         \includegraphics[width=\textwidth]{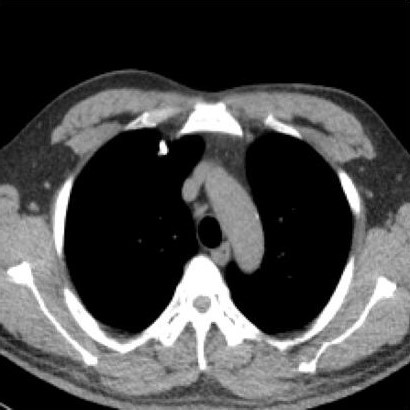}
         
     \end{subfigure}\hspace{0.05cm}%
     \hfill
     \begin{subfigure}[b]{0.12\textwidth}
         \centering
         
         \includegraphics[width=\textwidth]{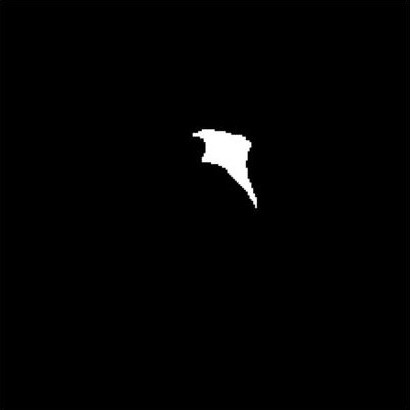}
         
     \end{subfigure}\hspace{0.05cm}%
     \hfill
     \begin{subfigure}[b]{0.12\textwidth}
         \centering
         
         \includegraphics[width=\textwidth]{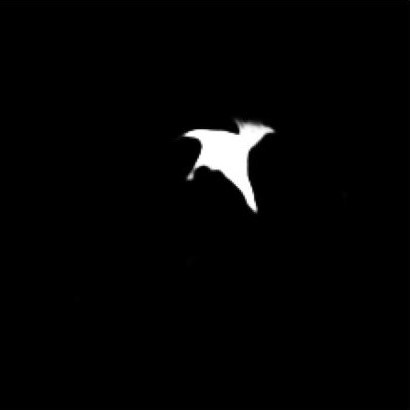}
         
     \end{subfigure}\hspace{0.05cm}%
     \hfill
     \begin{subfigure}[b]{0.12\textwidth}
         \centering
         
         \includegraphics[width=\textwidth]{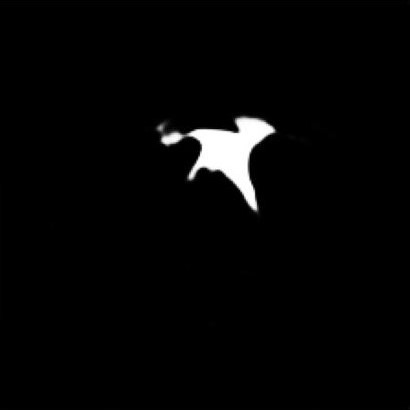}
    \end{subfigure}\hspace{0.05cm}%
     \hfill
     \begin{subfigure}[b]{0.12\textwidth}
         \centering
         
         \includegraphics[width=\textwidth]{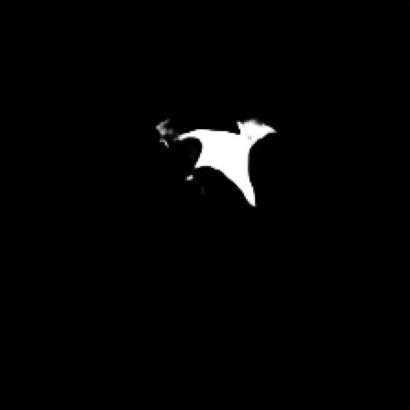}

     \end{subfigure}\hspace{0.05cm}%
     \hfill
     \begin{subfigure}[b]{0.12\textwidth}
         \centering
         
         \includegraphics[width=\textwidth]{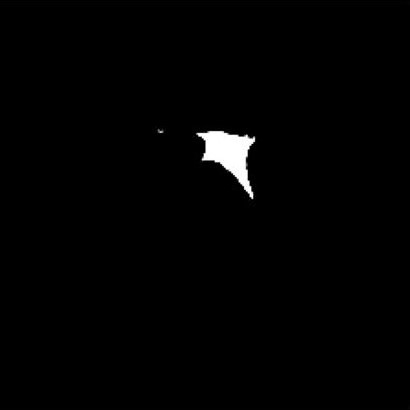}
         
     \end{subfigure}\hspace{0.05cm}%
     \hfill
     \begin{subfigure}[b]{0.12\textwidth}
         \centering
         
         \includegraphics[width=\textwidth]{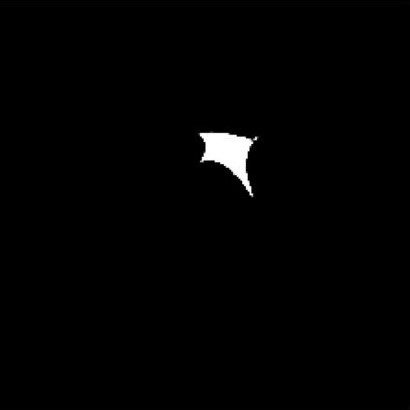}
         
     \end{subfigure}\hspace{0.05cm}%
     \hfill
     \begin{subfigure}[b]{0.12\textwidth}
         \centering
         
         \includegraphics[width=\textwidth]{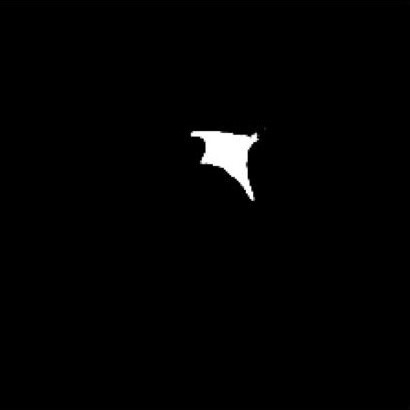}
         
     \end{subfigure}
        \caption{Visual comparison of distinct models.}
        \label{fig7}
\end{figure*}
\section{Evaluation} \label{Result}
In this section, we first compared the proposed method to the SOTA network. Then, in the subsection on the proposed method's effectiveness, we used different feature extractions, such as Resnet18 and the encoder part of the original Unet, to show the flexibility of our proposed architecture in using various feature extractions in the encoder part. These results showed that with almost the same structure in the encoder of Unet, Res\_Unet, and Res\_Unet++, the proposed method with different feature extraction performed better than those.
\subsection{Result}
Table ~\ref{table1} outlines the segmentation results of SOTA networks and our proposed model. As shown, because of various shapes of AM, just complicated network achieved acceptable results. Res\_Unet++ and Res\_Unet are two networks demonstrated inferior performance for all metrics. DCS, IoU, sensitivity, and Acc were obtained at 55.95\%, 39.93\%, 39.97\%, and 98.83\%, respectively, for Res\_Unet. Res\_Unet++ is the development of Res\_Unet, which performs better than Res\_Unet. All metrics were as follows: DSC = 59.80\%, IoU = 43.83, sensitivity = 43.92\%, and Acc 99.09\%.

Trans\_Unet is another SOTA network that was considered for comparison. Based on the transformer modules as encoder part of this network, all results were higher than Res\_Unet and Res\_Unet++. DCS, IoU, sensitivity, and ACC were achieved at 85.50\%, 75.98\%, 85.32\%, and 99.80, respectively, for Trans\_Unet. Attention\_Unet obtained 86.99\% in DSC, 78.28\% in IoU, 87.36\% in sensitivity, and 99.84\% Acc. 
Compared to all network, our proposed network performed better in DSC (87.69\%), IoU (78.81\%), and sensitivity (87.69\%) while it has significantly less parameters (6.7 M).

Fig.~\ref{fig7} shows the comparative analysis of several networks on our AM dataset. When the AM structures of interest have a small visual form, Trans\_Unet and the proposed network performed better in AM segmentation (second row of Fig. ~\ref{fig7}). The findings demonstrate the MHSA enhances the network's capacity to capture more contextual information, enabling it to distinguish the AM organ from the surrounding organs.
The results demonstrate the effectiveness of the MHSA in AM segmentation. Networks such as Unet, Res\_Unet, and Res\_Unet++, which do not utilize MHSA, struggle to segment AM. In contrast, attention\_Unet, Trans\_Unet, and the proposed, which incorporate attention, outperform the former.
\begin{table*}[h]
  \centering
  \caption{Results of AM segmentation for different methods}
  \resizebox{\textwidth}{!}{%
  \begin{tabular}{|c|c|c|c|c|c|}
    \hline
    \textbf{Model} & \textbf{DCS (\%)} & \textbf{IoU (\%)} & \textbf{Sensitivity (\%)} & \textbf{Acc (\%)}& \textbf{Parameters (M)}\\
    \hline
    Trans\_Unet & 85.50 & 75.98 & 85.32 &  99.80 & 88\\
    \hline
    Attention\_Unet & 86.99 & 78.28 & 87.36 &  \textbf{99.84} & 34.87\\
    \hline
    Unet & 63.04 & 47.05 & 47.11 & 98.21 & 31.03\\
    \hline
    Res\_Unet & 55.95 & 39.93 & 39.97 & 98.93& 13.05\\
    \hline
    Res\_Unet++ & 59.80 & 43.83 & 43.92 & 99.09& 14.48\\
    \hline
    Proposed network& \textbf{87.83} & \textbf{79.16} & \textbf{89.60} & 99.83 & \textbf{6.7}\\
    \hline
  \end{tabular}
  }
  \label{table1}
\end{table*}
\subsection{Effectiveness of proposed backbone}
\begin{figure}[t]
     \centering
     \begin{subfigure}[t]{0.27\textwidth}
         \centering
         \includegraphics[width=\textwidth]{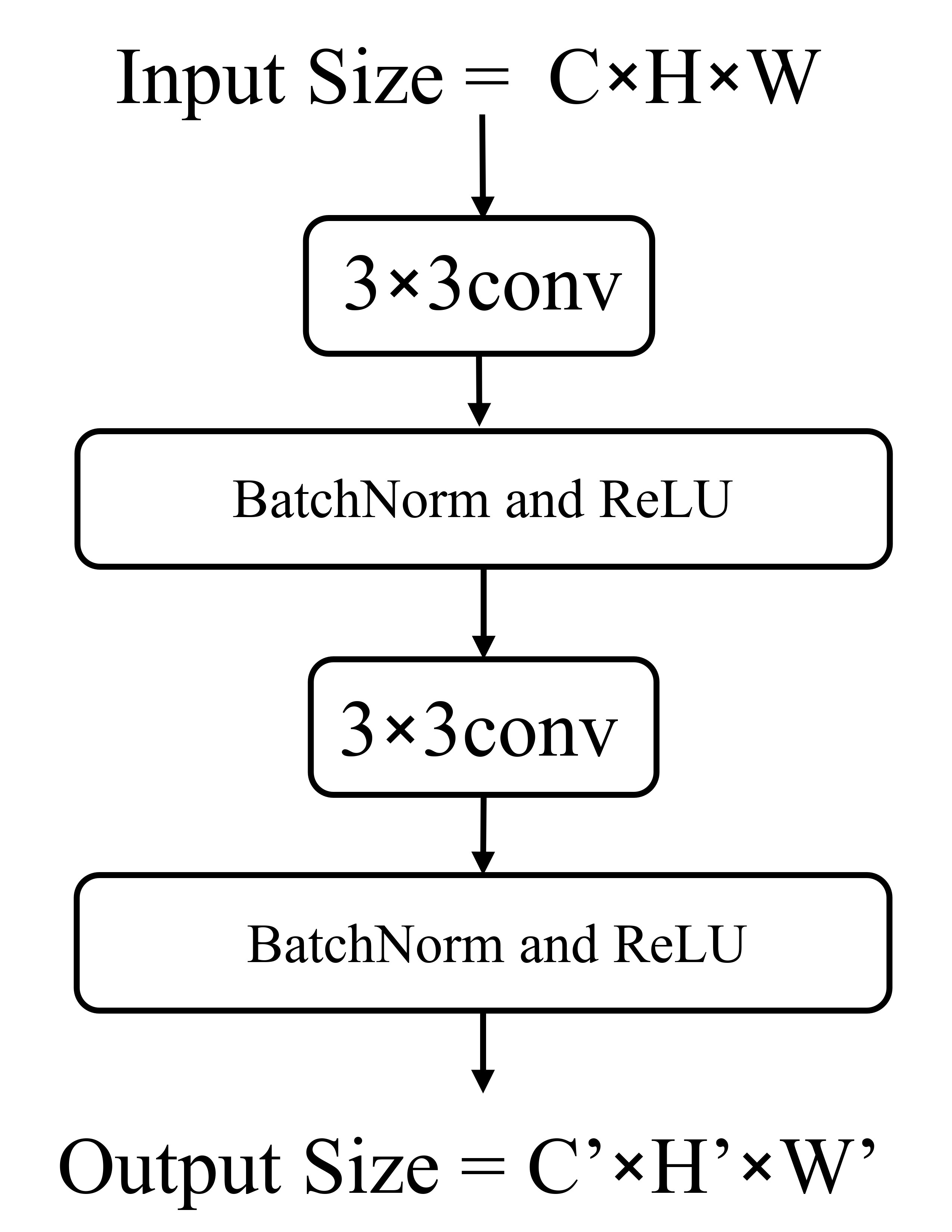}
         \caption{Consecutive Convolution Block (CCB)}
     \end{subfigure}\hspace{0.5cm}%
     \begin{subfigure}[t]{0.27\textwidth}
         \centering
         \includegraphics[width=\textwidth]{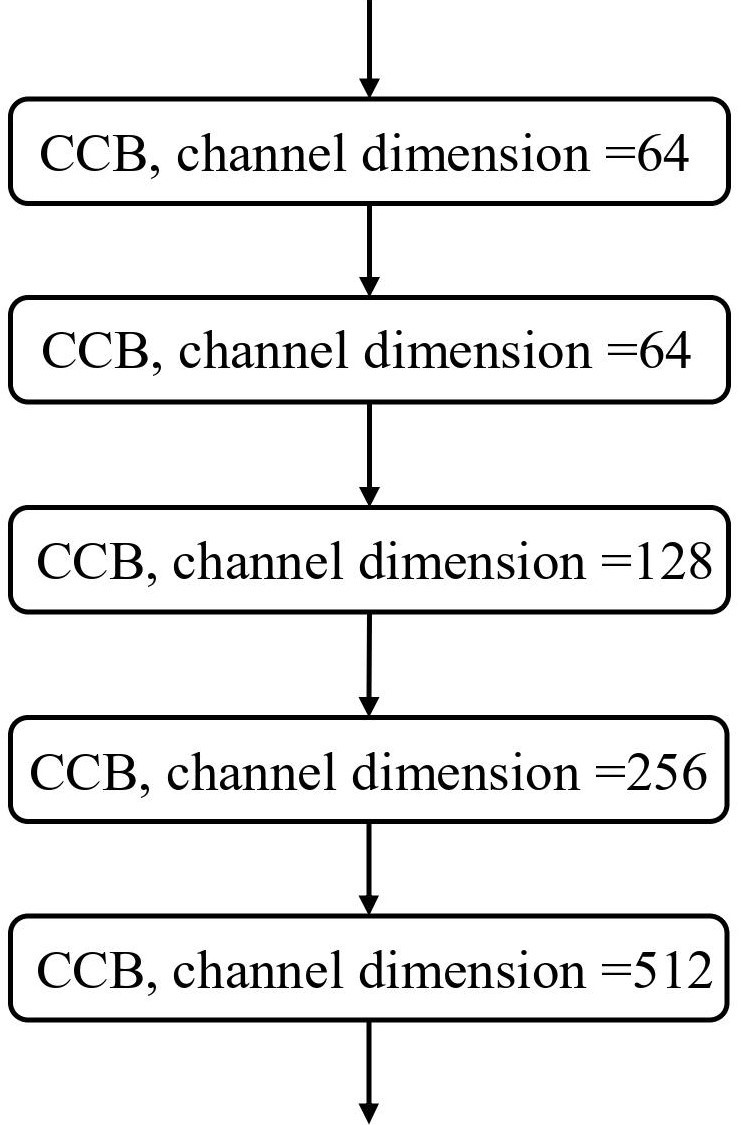}
         \caption{Encoder of original Unet \cite{ronneberger2015u}}
     \end{subfigure}
     \caption{(a) : Consecutive Convolution Block (CCB). The original block used in Unet \cite{ronneberger2015u} was designed without normalization layers. However, in this study we added batchnorm in to CCB. (b): We used this structure in Fig. ~\ref{figure1} as feature extractor.}
     \label{fig2_2}
\end{figure}

In Table ~\ref{table2_3}, we illustrated the effectiveness of the proposed Unet backbone by using Resnet18 and the encoder part of the original Unet. Proposed Unet (encoder = Resnet18) showed approximately 30\% higher performance in each metric while having higher parameters (15.2 M vs. 13.05 M and 14.48 M) than Res\_Unet and Res\_Unet++. The results of the proposed network illustrate that each metrics (DSC = 86.91\%, IoU = 77.75\%, sensitivity = 87.62\%, and Acc = 99.83\%) has competitive value compared to Attention\_Unet. In addition, a Consecutive Convolution Block (CCB)-based encoder of original Unet was designed. We used the same structure for the original Unet by adding normalization layers to be robust and efficient. As shown in Fig. ~\ref{fig2_2} (a), this convolution layer involves one 3×3 convolution, normalization layer and activation function, which are repeated twice. As shown in Fig. ~\ref{fig2_2} (b), encoder of Unet was designed in five stages with 64, 64, 128, 256, and 512 channels. Results of the proposed network (encoder = consecutive convolution block) demonstrated that our proposed structure is not limited to our proposed encoder. 
It showed DSC = 87.70\%, IoU = 78.83\%, sensitivity = 87.85\%, and Acc = 99.80\%, which is the second-best value behind the initially proposed network as shown Table ~\ref{table2_3}.

The results of this section demonstrates the effectiveness of the proposed DDWPP and CDWCC.

\begin{table*}[h]
  \centering
  \caption{Results of effectiveness the proposed backbone with different encoder.}
  \resizebox{\textwidth}{!}{%
  \begin{tabular}{|c|c|c|c|c|c|}
    \hline
    \textbf{Model (Proposed\_Unet)} & \textbf{DCS (\%)} & \textbf{IoU (\%)} & \textbf{Sensitivity (\%)} & \textbf{Acc (\%)}& \textbf{Parameters (M)}\\
    \hline
    Encoder: Expanding Convolution block & \textbf{87.83} & \textbf{79.16} & \textbf{89.60} & \textbf{99.83} & \textbf{6.7}\\
    \hline
    Encoder: Resnet18 & 86.91 & 77.75 & 87.62 & 99.83& 15.2\\
    \hline
    Encoder: Consecutive convolution block & 87.70 & 78.83 & 87.85 & 99.80 &8.8\\
    \hline
  \end{tabular}
  }
  \label{table2_3}
\end{table*}

\section{Conclusion}\label{Conclusion}
To automatically detect Anterior Mediastinum (AM) in the Computed Comography (CT) scans, we proposed an architecture based on the U-shaped structure. For our proposed network, an expanded convolution block and W-Multi-Head Self-Attention (W-MHSA) were utilized as feature extraction. Channel Depth Wise Cross Correlation Attention (CDWCC) was introduced in decoder of the proposed network to find similarities in feature maps from the encoder and decoder. In addition, Dilated Depth-Wise Parallel Path connection (DDWPP) has been proposed as connection between encoder and decoder of the proposed U-shaped network in order to maintain long range dependencies. A dataset of 200 AM patients was used to justify the effectiveness of the proposed network. According to the results, the proposed network obtained 87.83\% in DSC, 79.16\% in IoU, and 89.60\% in Sensitivity, which are higher than those of SOTA networks, while it has significantly fewer parameters (6.7 M).

For future work, we anticipate the following clinical utilities for the proposed model for the automatic segmentation of AM in this study: (i) automatic detection and classification of AML in chest CT, including LDCT; (ii) distribution analysis of anterior mediastinal volume according to clinical factors such as age and sex; and (iii) educational purposes for medical students, healthcare professionals, and the general public.
\newline
\newline

\bibliographystyle{IEEEtran} 

\bibliography{references.bib}

\end{document}